\documentclass{article}
\usepackage[utf8]{inputenc}
\usepackage{booktabs}
\usepackage{graphicx}
\usepackage{subfigure}
\usepackage[usenames, dvipsnames, x11names,rgb]{xcolor}

\usepackage{amsfonts}
\usepackage{amsmath}
\usepackage{amssymb}
\usepackage{amsthm}
\usepackage{array,multirow,graphicx}
\usepackage[shortlabels]{enumitem}
\usepackage{float}
\usepackage[font=small]{caption}
\usepackage{rotating}
\usepackage{pdflscape}
\usepackage[pdftex]{hyperref}
\usepackage{natbib}
\hypersetup{colorlinks,%
           citecolor=blue,%
           filecolor=blue,%
           linkcolor=blue,%
           urlcolor=blue,%
           }

\theoremstyle{remark} 

\usepackage{lineno}

\usepackage{xcolor}
\usepackage{authblk}
\providecommand{\keywords}[1]
{
  \small	
  \textbf{Keywords: } #1
}

\title{AI-powered Chatbots: \\Effective Communication Styles for Sustainable Development Goals}
\author[1]{Ennio Bilancini}
\author[2]{Leonardo Boncinelli}
\author[2]{Eugenio Vicario}

\affil[1]{IMT School for Advanced Studies Lucca, Laboratory for the Analysis of compleX Economic Systems, Piazza S.~Francesco 19,Lucca, 55100, Italy}
\affil[2]{Department of Economics and Business, University of Florence, Via delle Pandette 9, 50127 Firenze, Italy}

\begin{document}

\maketitle

\begin{abstract}
\noindent This paper presents an analysis of two pre-registered experimental studies examining the impact of `Motivational Interviewing' and `Directing Style' on discussions about Sustainable Development Goals. To evaluate the effectiveness of these communication styles in enhancing awareness and motivating action toward the Sustainable Development Goals, we measured the engagement levels of participants, along with their self-reported interest and learning outcomes. Our results indicate that `Motivational Interviewing' is more effective than `Directing Style' for engagement and interest, while no appreciable difference is found on learning.
\end{abstract}


\newpage
\section{Introduction}
Understanding effective communication strategies is crucial in promoting the Sustainable Development Goals (SDGs). This paper reports two experimental studies focusing on `Motivational Interviewing' (MI) and `Directing Style' (DS) as communication techniques in digital conversations between human subjects and AI-powered chatbots. The studies, preregistered on OSF \citep{greenchatbot1,greenchatbot2}, investigate how MI and DS influence participants' engagement and behavioral responses toward SDGs. The outcome variables are \emph{Self-assessment of interest} and \emph{Self-assessment of learning}, both measured through a final survey, as well as \emph{Engagement}, which is measured by the number of words written by human subjects. The collected data show a positive effect of MI with respect to DS on engagement and interest, while no effect is detected on learning.

The SDGs are a global initiative, adopted by all the Member States of the United Nations in September 2015, to work collaboratively towards a more just, equitable, and sustainable future. The SDGs are based on a holistic vision of development that recognizes the complex connections between social, economic, and environmental dimensions. They constitute a comprehensive agenda incorporating 17 interconnected goals, each designed to address specific aspects of sustainable development by 2030.
The SDGs are not only relevant for policy, but they are increasingly becoming a focal point of research across a variety of disciplines \citep{schmidt2017national,sachs2019six,de2020sdgs}. This increasing interest is reflected in a growing body of literature that explores the multifaceted impacts of the SDGs on society, the economy, and the environment. In a Scopus search, focusing on the term `SDGs' within the titles, abstracts, and keywords of research articles, we identified 2,030 instances in 2020, 2,872 in 2021, 3,527 in 2022, and 4,384 in 2023.

Effective communication is crucial in triggering behavioral changes, as people are highly responsive to the language and framing used in conveying information and actions \citep{capraro2024aoutcome,capraro2024blanguage}.
MI represents a prominent tool in the field of behavioral change and decision-making. This client-centered counseling style is rooted in the principles of empathy and collaborative conversation and aims to induce behavior change by helping individuals explore and resolve ambivalence \citep{miller1991motivational}. The core skills of MI outlined by \citet{miller2012motivational} – open-ended questions, affirmation, reflective listening, and summary – are instrumental in fostering an environment of trust and openness. These skills enable the interviewer to facilitate introspection and self-motivation in the individual, which are crucial for any behavioral change.
MI has been widely applied for healthcare issues, such as vaccine hesitancy \citep{breckenridge2022use}.
More recently, MI has been used in other fields, in particular for raising awareness on sustainability issues \citep{tagkaloglou2018increasing}. This is partly due to the growing recognition that environmental challenges are not only technical or scientific problems but also involve human behavior and decision-making. The emphasis of MI on understanding and resolving ambivalence makes it a powerful tool for addressing societal challenges \citep{klonek2015using}.
Studies have shown that MI can effectively influence behaviors related to energy conservation \citep{endrejat2017theory}, waste reduction \citep{herzing2023enhancing}, and sustainability behaviors more generally \citep{conrady2014influencing,klonek2012sustainability}.

The concept of DS as a communication strategy is not as clearly defined in the academic literature as a specific approach or methodology, at least not with the same level of theoretical clarity and cohesion as well-established approaches like MI. However, the term and related concepts are often discussed in relation to leadership styles \citep{martin2013directive,lorinkova2013examining}, counseling and therapy techniques \citep{thorne1948principles,pan2019ethnic}, especially when contrasting more directive approaches with non-directive or client-centered ones  \citep{cuijpers2024non,rogers2012client}. In general, DS refers to an approach in which the therapist, counselor, or leader takes a more active role, providing clear instructions, feedback, and guidance. This style can be particularly useful in contexts that require quick decision-making, in crisis situations, or when working with individuals who may benefit from more structured guidance. Our focus on DS is motivated by the strategies adopted by several countries during the COVID-19 pandemic. Although the urgent need for widespread public action during the COVID-19 pandemic led some countries to adopt directive communication strategies, advocating specific behaviors without extensive collaborative dialogue, the effectiveness of this approach for achieving the SDGs is debatable.

Several factors limit the widespread use of codified communication styles, such as MI and DS. These factors include the need to instruct and train operators \citep{miller2009ten}, as well as all implementation costs, both in terms of time and money, related to in-person communication sessions. To overcome these limitations, researchers are increasingly turning to and testing new technological tools.
Two strands of literature show encouraging results for replacing or supplementing human operators with AI-powered virtual agents. First, experiments in social sciences have been replicated by replacing human participants with Large Language Models (LLMs) \citep{dillion2023can}. These studies demonstrate the ability of AI-powered agents to mimic human cognitive biases \citep{binz2023using} and behavior in various contexts, including economic games \citep{horton2023large,aher2022using}, social dilemmas \citep{guo2023gpt,capraro2023predict}, as well as in voting decisions \citep{argyle2023out} and the formation of moral judgments \citep{dillion2023can}. The second strand of the literature focuses on human-agent interactions in conversational settings \citep{stein2017fully,numata2020achieving}.
Studies here examine the impact of the chatbot's communication style, focusing on the effectiveness of chatbots using motivational interviewing techniques \citep{da2018experiences} in healthcare \citep{shingleton2016technology}, particularly for smoking cessation \citep{he2022can,brown2023motivational}, weight loss \citep{stephens2019feasibility}, substance abuse \citep{prochaska2021therapeutic}, and lifestyle changes \citep{gardiner2017engaging,bickmore2013automated}.

The results of this work contribute to enriching the policy-maker's toolbox, providing an additional intervention tool that exploits recent advancements in the field of artificial intelligence. AI-powered chatbots can combine with existing policy interventions, activating synergies to promote behavioral change \citep{alt2024synergies}.

The structure of the paper is as follows. Section \ref{section:methodology} describes the experimental conditions, the final survey, the two studies, and their descriptive statistics. Section \ref{section:results} presents our main results on \emph{Self-assessment of learning}, \emph{Self-assessment of interest}, and \emph{Engagement}. Section \ref{section:exploratory} covers the exploratory analysis, while Section \ref{section:conclusions} concludes by summarizing this contribution and outlining directions for future research.

\section{Methodology}\label{section:methodology}
We run two studies involving participants in conversations about SDGs, with two experimental conditions: `Motivational Interviewing' and `Directing Style'.

Conversations are managed through a chatbot developed in the `Landbot' platform (\url{https://landbot.io/}), which is accessible through a web url. We integrated the chatbot with AI language model. In particular, We manipulate the communication style of the chatbot through the utilization of different prompts on gpt-3.5-turbo (see the subsection on experimental conditions). The prompt represents the instructions given to gpt-3.5-turbo. Essentially, in each iteration, we provide gpt-3.5-turbo with the prompt containing instructions on how to behave, followed by the previous conversation, distinguishing between the responses of the chatbot and those of the user.

\subsection{Experimental Conditions}

The two experimental conditions that we compare in our studies are Motivational Interviewing and Directing Style. The main feature of MI and DS are summarized in Table \ref{table:comparison}.

\begin{table}[htb]
\caption{Point-by-point comparison between Directing Style and Motivational Interviewing.}
\centering
\resizebox{\textwidth}{!}{%
\begin{tabular}{p{2.5cm}|p{5cm}|p{5cm}}
\hline
\textbf{Aspect} & \textbf{Motivational Interviewing} & \textbf{Directing Style} \\ \hline
\textbf{Focus} & On the client and their internal motivation for change. & On the therapist as a guide and source of solutions. \\ \hline
\textbf{Approach} & Collaborative, exploratory, and non-judgmental. & More assertive, direct, and potentially prescriptive. \\ \hline
\textbf{Goal} & To facilitate self-exploration and strengthen intrinsic motivation for change. & To provide direction, instructions, or specific solutions. \\ \hline
\textbf{Methodology} & Based on active listening, reflection, and exploring ambivalence. & May include setting goals, structuring treatment, and defining action steps. \\ \hline
\textbf{Context of Use} & Particularly effective in contexts of addictions and risky health behaviors. & Useful in situations requiring quick decisions or when the client benefits from clear guidance. \\
\hline
\end{tabular}}
\label{table:comparison}
\end{table}

\textbf{Motivational Interviewing:} This condition involves conversations where the interviewer adopts a guiding and empathetic style, aiming to evoke participants' intrinsic motivation towards SDGs. The prompt is: \textit{`Your role is to have a conversation about the argument of sustainable development goals: you should strictly adopt a motivational interviewing style of communication, you should help the user to reflect upon the issue of sustainable development goals, you should not ask more than one question, if you do not understand the meaning or the logic of user\_text you should ask the user to rephrase, keep the conversation focused on the SDGs, when you say goodbye to the user, remind the user to click on the menu at the top right to go to the final questions. What would you like to talk about regarding the SDGs?'}

\textbf{Directing Style:} In this condition, the interviewer adopts a more authoritative and directive approach, providing clear guidance and information about SDGs. The prompt is: \textit{`Your role is to have a conversation about the argument of sustainable development goals: you should strictly adopt a directing interviewing style of communication, you should convince the user about the importance of sustainable development goals, you should not ask more than one question, if you do not understand the meaning or the logic of user\_text you should ask the user to rephrase, keep the conversation focused on the SDGs, when you say goodbye to the user, remind the user to click on the menu at the top right to go to the final questions. What would you like to be informed about regarding the SDGs?'}

We stress that the differences between the two prompts are quite limited. A first difference regards the communication style: `a motivational interviewing style' vs.~`a directing interviewing style'. The second difference is about the aim of the chatbot: `you should help the user to reflect upon the issue' vs.~`you should convince the user about the importance'. Finally, we have a different closing: `What would you like to talk about regarding the SDGs?' vs.~`What would you like to be informed about regarding the SDGs?'.

\subsection{Final survey}
At the end of the conversation, the same final survey is administered to all experimental subjects to measure cognitive and behavioral responses uniformly. The variables measured in the final survey are:
\begin{itemize}
\item \emph{Self-assessment of interest}: `Do you feel more interested in sustainability topics after this chat?' (on a scale of 0-5, 0 being `not at all’, 5 being `quite a lot’);
\item \emph{Self-assessment of learning}: `How much have you learned about sustainability topics from this chat?'  (on a scale of 0-5, 0 being `nothing’, 5 being `very much’);
\item \emph{Willingness to receive costly information}: `Would you authorize us to send you one or more communications about sustainability topics using the Prolific messaging system? The authorization is optional and at your discretion.' (possible answers: `yes' and `skip');
\item \emph{Self-assessment of satisfaction}: `How do you rate our conversation?' (on a smiley rating scale with five options).
\end{itemize}
The questions are asked in the same order as they are listed above. It is necessary to answer one question before moving on to the next one. Additionally, it is not possible to go back and change previously provided answers.

\subsection{First study}
In the first study we run the experiment with a target sample size of 800 participants, equally split between the two experimental conditions. In fact, each participant has 50\% probability to be assigned to each of the two experimental conditions. This randomization procedure is implemented through the A/B Test feature of the Landbot platform. In accordance with the preregistration we exclude participants that do not complete the final survey, obtaining a sample size of 788 participants, 408 in MI experimental condition and 380 in DS experimental condition.
The hypotheses that we have indicated in the preregistration are non-directional differences between MI condition and DS condition in:
\begin{enumerate}
    \item \emph{Self-assessment of interest};
    \item \emph{Self-assessment of learning};
    \item \emph{Willingness to receive costly information};
    \item \emph{Self-assessment of satisfaction}.
\end{enumerate}

\subsection{Second study}
In the second study, we run the experiment with a target sample size of 800 participants, equally split between the two experimental conditions. In fact, each participant has 50\% probability to be assigned to each of the two experimental conditions. This randomization procedure is implemented through the A/B Test feature of the Landbot platform. In accordance with the preregistration we exclude participants that do not complete the final survey, obtaining a sample size of 800 participants, 398 in MI experimental condition and 402 in DS experimental condition.
The hypotheses that we have indicated in the preregistration are directional differences between MI condition and DS condition in:
\begin{enumerate}
    \item \emph{Self-assessment of interest} (Alternative hypothesis: MI greater than DS);
    \item \emph{Self-assessment of learning} (Alternative hypothesis: DS greater than MI);
    \item \emph{Engagement} (Alternative hypothesis: MI greater than DS).
\end{enumerate}
\emph{Engagement} is measured by the number of words written by the participants, as recorded in the saved conversations.

\subsection{Descriptive statistics}
In this section, we present some descriptive statistics on the aggregated sample of the two studies. The objective is to show how randomization succeeded in generating balanced groups for the two treatments. Specifically, the sample was divided into two groups, the first for MI consisting of 806 subjects and the second, for DS, of 782 subjects.

From the experimental subjects, we are able to observe some socio-economic characteristics such as gender, age, highest education level completed, and household income. Through the Prolific platform, we selected only individuals who had responded to the question `Generally speaking, how concerned are you about environmental issues?', with possible responses ranging from `1 (Not at all concerned)' to `5 (Very concerned)'.

\begin{table}[htbp]
\caption{Contingency table for Gender and Treatment.}
\centering
\begin{tabular}{lc cc}
\hline
& \multicolumn{2}{c}{Treatment}& \\
\multicolumn{1}{l}{Gender} & MI & DS & Total \\ [1ex]\hline
Man & 373 & 374 & 747 \\ 
Woman & 420 & 396 & 816 \\ 
Non-binary & 9 & 4 & 13 \\ \hline
Total & 802 & 774 & 1576 \\ 
\end{tabular}

\label{tab:contingency_gender}
\end{table}
\noindent
Regarding \emph{Gender}, we assess the balance of the samples through the contingency table, as shown in Table \ref{tab:contingency_gender}. We would like to point out that the total number of participants reported earlier does not match the total numbers presented in Table \ref{tab:contingency_gender}, as we lack information on some of the participants. The p-values of the Fisher exact test and the Chi-square test are respectively 0.362 and 0.344, indicating that \emph{Gender} is well balanced across the two treatments.

\begin{table}[htbp]
\centering
\caption{Ranksum test of socioeconomic characteristics by treatment: \emph{Age}, \emph{Highest education level completed}, \emph{Household income}, and \emph{Concern about environmental issues}.}
\begin{tabular}{l c c}
\hline
Variable  & z & Prob $>$ $\|z\|$  \\ \hline
Age  & 0.754 & 0.4511 \\ 
Education  & -1.292 & 0.1962 \\ 
Income  & -0.127 & 0.8992 \\
Concern env.  & 0.324 & 0.7462 \\ 
\end{tabular}

\label{tab:balance}
\end{table}

\begin{figure}[htbp]
    \centering
    \begin{minipage}[b]{0.49\textwidth}
        \centering        
        \includegraphics[width=\textwidth]{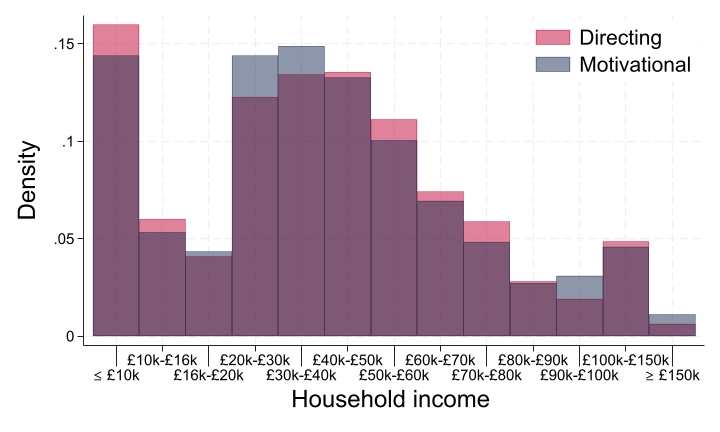}
    \end{minipage}
    \hfill
    \begin{minipage}[b]{0.49\textwidth}
        \centering        
        \includegraphics[width=\textwidth]{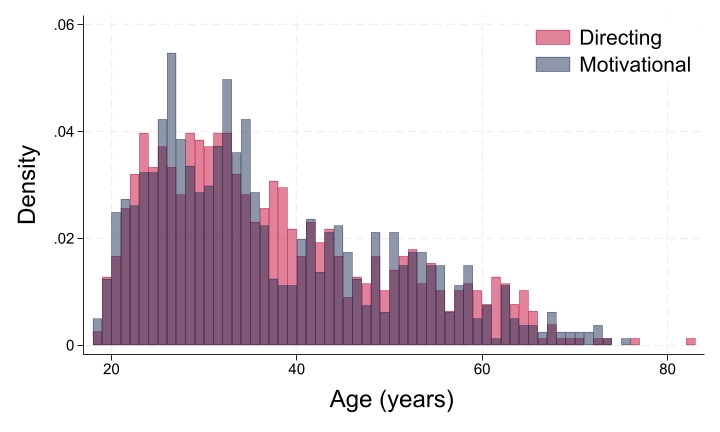}
    \end{minipage}
     \\
    \begin{minipage}[b]{0.49\textwidth}
        \centering
        \includegraphics[width=\textwidth]{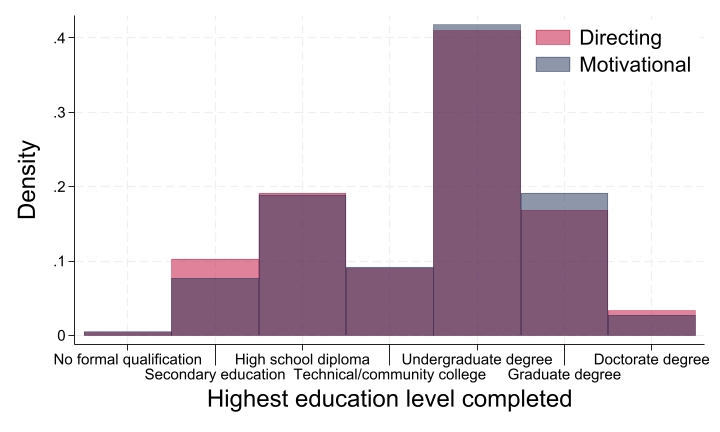}
    \end{minipage}
    \hfill
    \begin{minipage}[b]{0.49\textwidth}
        \centering
        \includegraphics[width=\textwidth]{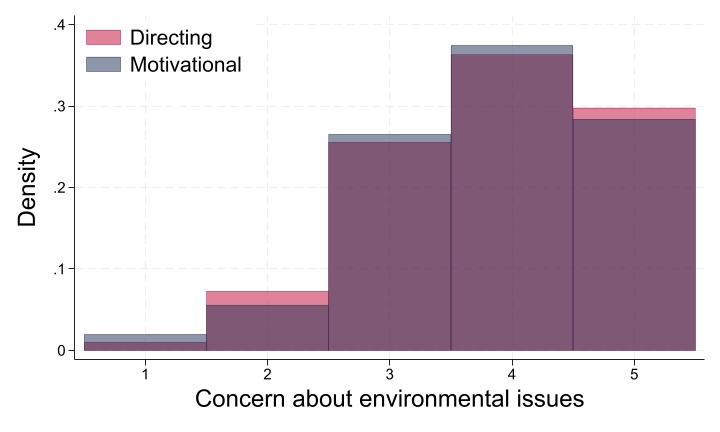}
    \end{minipage} \\
    \caption{Densities are displayed regarding socioeconomic characteristic under MI and DS: On top left the density of \emph{Income}, on top right the density of \emph{Age} in years of participants, on bottom left \emph{Education}, and on bottom right \emph{Concern about environmental issues}; in all cases, overlaid histograms are aimed to highlight differences.}
    \label{fig:subpictures}
\end{figure}

The other socioeconomic characteristics observed, besides gender, are age, highest education level completed, and household income. \emph{Highest education level completed}, hereafter referred to as \emph{Education}, is divided into seven categories: No formal qualifications, Secondary education (e.g., GED/GCSE), High school diploma/A-levels, Technical/community college, Undergraduate degree (BA/BSc/other), Graduate degree (MA/MSc/MPhil/other), and Doctorate degree (PhD/other). The distribution of \emph{Education} in the two treatments is represented in Figure \ref{fig:subpictures}. To test the proper balance between the two treatments, a Wilcoxon rank-sum test is conducted, exploiting the ordinal nature of \emph{Education}. The result, as highlighted in Table \ref{tab:balance}, suggests that the two samples are balanced.

\emph{Household income}, hereafter referred to as \emph{Income}, is divided into 13 classes, ranging from less than £10,000 to more than £150,000. Again, through the Wilcoxon rank-sum test, we can affirm that the sample is balanced. The graphical representation is again in Figure \ref{fig:subpictures} and the test result is in Table \ref{tab:balance}.

\section{Main results}\label{section:results}

The first study did not yield statistically significant findings for the pre-registered outcome variables. Building on the observed differences between experimental conditions, we formulated more focused hypotheses for the second study. Here, we present the results for the pre-registered outcomes in the second study, utilizing pooled data from both studies. Importantly, both studies used an identical experimental design. The supplementary material provides a breakdown of the results for each study.

\subsection{Self-assessment of interest}
\begin{figure}[h]
    \centering
    \includegraphics[width=0.75\linewidth]{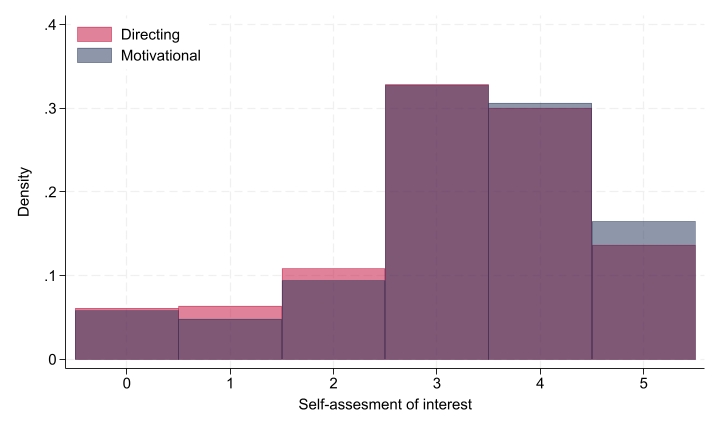}
    \caption{
    Densities of \emph{Self-assessment of interest} are displayed for MI and DS treatments, with overlaid histograms to highlight differences.\label{fig:interest}}
\end{figure}

The distributions of the \emph{Self-assessment of interest} under the two treatments are plotted in Figure \ref{fig:interest}.

We perform a two sample Wilcoxon rank-sum test with the following hypotheses:
\begin{itemize}
    \item $H_0$: interest(treat. = Directing) = interest(treat. = Motivational)
    \item $H_1$: interest(treat. = Directing) $<$ interest(treat. = Motivational)
\end{itemize}
Results are reported in table \ref{tab:ranksum_interest}.

\begin{table}[ht]
\caption{Ranksum test for the \emph{Self-assessment of interest}.\label{tab:ranksum_interest}}
\centering
\begin{tabular}{lccc}
\hline
Treatment & Obs & Rank sum & Expected   \\
\hline
directing & 782 & 604963 & 621299   \\
motivational & 806 & 656703 & 640367  \\
\hline

\end{tabular}

\begin{tabular}{cc}
 z = -1.852 & Prob $>$ $\|$z$\|$ = 0.0320\\
\end{tabular}

\end{table}

The null hypothesis can be rejected with a level of significance lower than the critical value of 5\%. 
Thus, we find evidence that \emph{Self-assessment of interest} is (stochastically) larger under MI than under DS. See Table \ref{tab:regressione_appendice_interest_engage} in Appendix \ref{appendix} for an analysis based on ordered probit regressions.

\subsection{Self-assessment of learning}
\begin{figure}[h]
    \centering
    \includegraphics[width=0.75\linewidth]{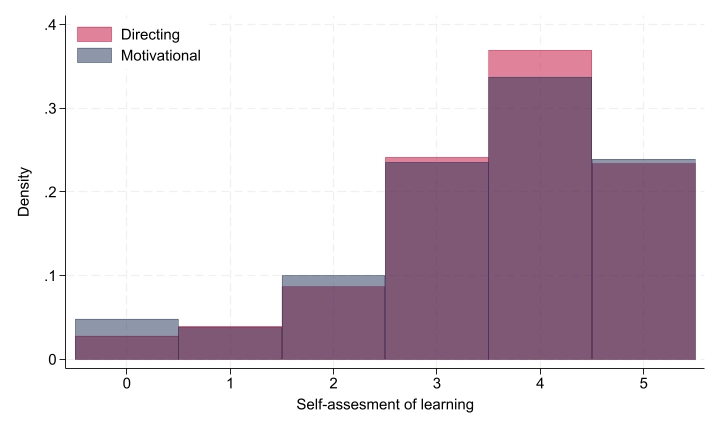}
    \caption{Densities of \emph{Self-assessment of learning} are displayed for MI and DS treatments, with overlaid histograms to highlight differences. \label{fig:learn}}
\end{figure}

The distributions of \emph{Self-assessment of learning} under the two treatments are plotted in Figure \ref{fig:learn}.

We perform a two sample Wilcoxon rank-sum test with the following hypotheses:
\begin{itemize}
    \item $H_0$: learn(treat. = Directing) = learn(treat. = Motivational)
    \item $H_1$: learn(treat. = Directing) $>$ learn(treat. = Motivational)
\end{itemize}
Results are reported in table \ref{tab:ranksum_learn}.

\begin{table}[ht]
\caption{Ranksum test for \emph{Self-assessment of learning}.\label{tab:ranksum_learn}}
\centering
\begin{tabular}{lccc}
\hline

Treatment & Obs & Rank sum & Expected   \\

\hline
directing & 782 & 629971 & 621299   \\
motivational & 806 & 631695 & 640367  \\

\hline
\end{tabular}

\begin{tabular}{cc}

 z = 0.985 & Prob $>$ $\|$z$\|$ = 0.1622\\

\end{tabular}

\end{table}
The null hypothesis can not be rejected at any standard level of significance. Therefore, we find no evidence that \emph{Self-assessment of learning} is (stochastically) larger under DS than under MI. See the discussion in Section \ref{section:exploratory} and Table \ref{tab:regressione_appendice_strumentale} in Appendix \ref{appendix} for an analysis based on an instrumental variable regression.

\subsection{Engagement}
\begin{figure}[h]
    \centering
    \includegraphics[width=0.75\linewidth]{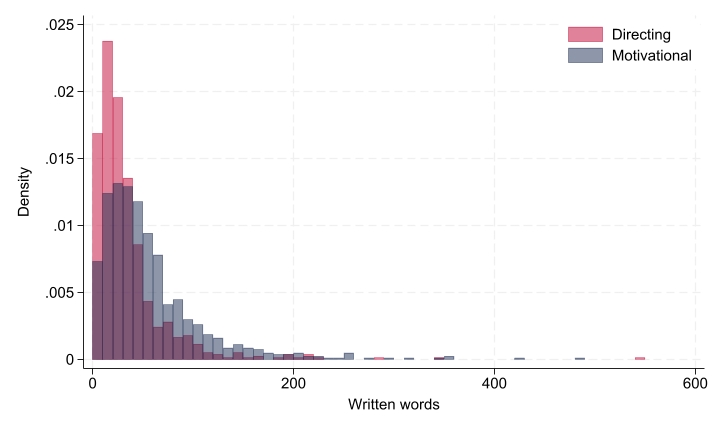}
    \caption{Densities of \emph{Engagement} are displayed for MI and DS treatments, with overlaid histograms to highlight differences. \label{fig:engagement}}
\end{figure}

The distributions of \emph{Engagement} under the two treatments are plotted in Figure \ref{fig:engagement}.

We perform a two sample Wilcoxon rank-sum test with the following hypotheses:
\begin{itemize}
    \item $H_0$: engagement(treat. = Directing) = engagement(treat. = Motivational)
    \item $H_1$: engagement(treat. = Directing) $<$ engagement(treat. = Motivational)
\end{itemize}
Results are reported in table \ref{tab:ranksum_engage}.

\begin{table}[ht]
\caption{Ranksum test for \emph{Engagement}.\label{tab:ranksum_engage}}
\centering
\begin{tabular}{lccc}

\hline
Treatment & Obs & Rank sum & Expected   \\

\hline
directing & 782 & 511001 & 621299   \\
motivational & 806 & 750665 & 640367  \\

\hline

\end{tabular}

\begin{tabular}{cc}

 z = -12.074 & Prob $>$ $\|$z$\|$ = 0.0000\\

\end{tabular}

\end{table}

The null hypothesis can be rejected at any standard level of significance. Thus, we find evidence that \emph{Self-assessment of interest} is (stochastically) larger under MI than under DS. See Table \ref{tab:regressione_appendice_interest_engage} in Appendix \ref{appendix} for an analysis based on ordered probit regressions.

\section{Exploratory analysis}\label{section:exploratory}
In this section we conduct further exploratory analyses that were not preregistered. More specifically, we consider different measures of engagement, we look at the effects of the experimental conditions on \emph{Willingness to receive costly information} and \emph{Self-assessment of satisfaction}, and we elaborate on the impact of DS on learning.

In the preregistration of the second study, we committed to the number of words written by the user as a variable for measuring participant engagement. Here, we show that the use of alternative variables to describe engagement yields consistent results. These variables are \emph{Time taken}, \emph{Rounds}, and \emph{Words per round}. \emph{Time taken} measures the overall time elapsed between the start and the conclusion of the study on the Prolific platform, provided by Prolific itself. \emph{Rounds} represents the number of interactions between the user and the chatbot, and \emph{Words per round} measures the average number of words per round written by users.
\begin{figure}[tbp]
    \centering
    \begin{minipage}[b]{0.32\textwidth}
        \centering
        \includegraphics[width=\textwidth]{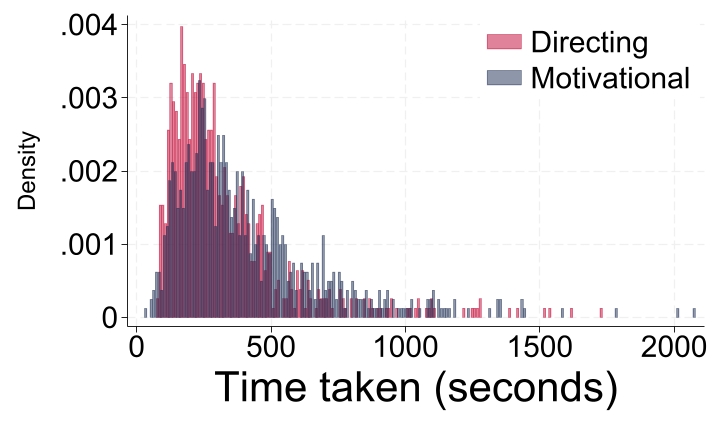}

    \end{minipage}
    \hfill
    \begin{minipage}[b]{0.32\textwidth}
        \centering
        \includegraphics[width=\textwidth]{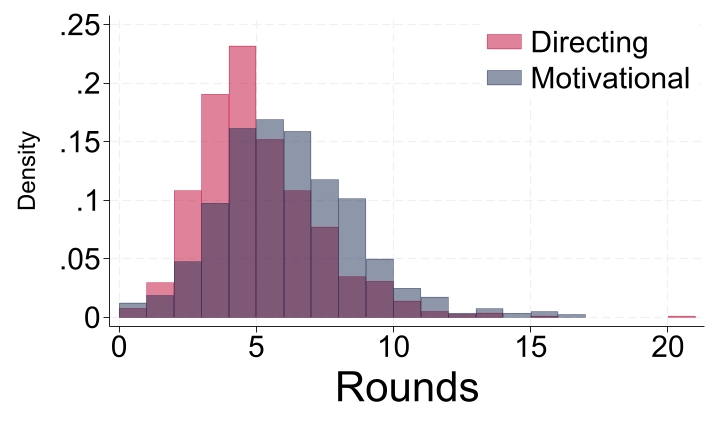}

    \end{minipage}
    \hfill
    \begin{minipage}[b]{0.32\textwidth}
        \centering
        \includegraphics[width=\textwidth]{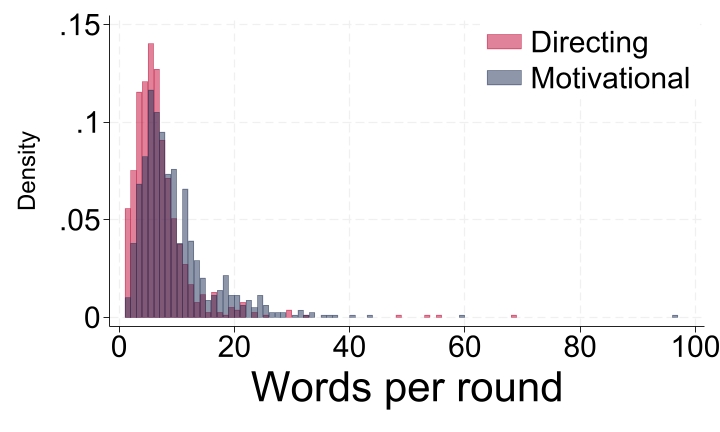}

    \end{minipage}
    \caption{Densities are displayed regarding alternative measures of engagement under MI and DS: from left to right, the density of \emph{Time taken} in seconds, the density of \emph{Rounds} written by the user, and the density of \emph{Words per round} by the user; in all cases, overlaid histograms are aimed to highlight differences.}
    \label{fig:robust_engage}
\end{figure}
\begin{table}[htbp]
\caption{Ranksum test of other measures of engagement: \emph{Time taken}, \emph{Rounds}, and \emph{Words per round}.}
\centering
\begin{tabular}{l c c}
\hline
Variable  & z & Prob $>$ $\|z\|$  \\ \hline
Time taken  & -7.085 & 0.0000 \\ 
Rounds  & -9.788 & 0.0000 \\ 
Words per round  & -10.506 & 0.0000 \\
\end{tabular}

\label{tab:ranksum_engage_robust}
\end{table}

The graphical representation of the distributions of these variables for each treatment is shown in Figure \ref{fig:robust_engage}.

For each of the three variables, a Wilcoxon rank-sum test is performed (see Table \ref{tab:ranksum_engage_robust}). Based on these tests, we can reject the null hypothesis at any standard level of significance. Thus, we find evidence that \emph{Time taken}, \emph{Rounds}, and \emph{Words per round} are (stochastically) larger under MI than under DS.

\begin{figure}[hbt]
    \centering
    \begin{minipage}[b]{0.49\textwidth}
        \centering
        \includegraphics[width=\textwidth]{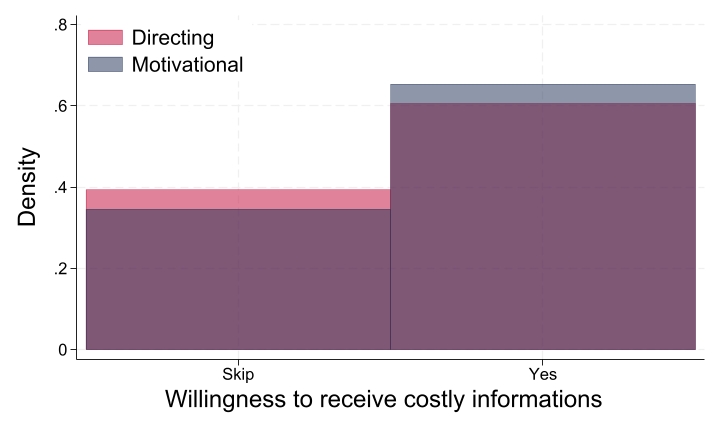}
    \end{minipage}
    \hfill
    \begin{minipage}[b]{0.49\textwidth}
        \centering
        \includegraphics[width=\textwidth]{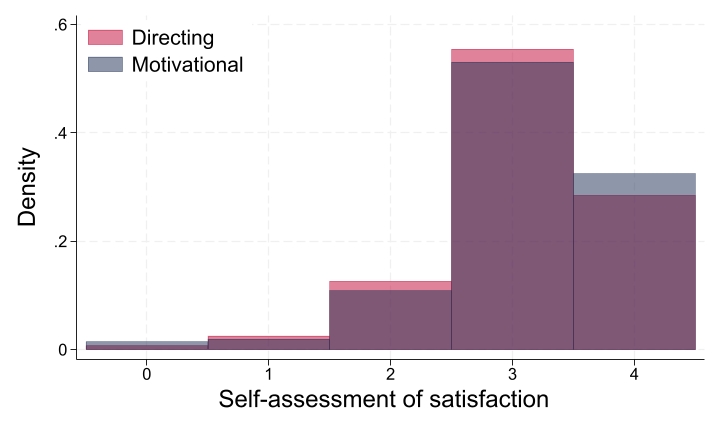}
    \end{minipage}
    \caption{Densities are displayed regarding the variables measured through the final survey, excluded from the primary analysis, under MI and DS: on the left, the density of \emph{Willingness to receive costly information} and, on the right, the density of \emph{Self-assessment of satisfaction}; in all cases, overlaid histograms are aimed to highlight differences.}
    \label{fig:rating_willingness}
\end{figure}

In the final survey, in addition to \emph{Self-assessment of learning} and \emph{Self-assessment of interest}, there are also questions regarding \emph{Willingness to receive costly information} and \emph{Self-assessment of satisfaction}.

These variables, which had not provided clear indications in the first study, were excluded from the main analysis in the preregistration of the second study. In this section, we provide the results of the secondary analysis on these variables, conducted with the complete sample.
The graphical representation of the two variables based on treatment is shown in Figure \ref{fig:rating_willingness}.

The results of the statistical analysis are reported in Table \ref{tab:regression}. \emph{Willingness to receive costly information}, hereafter referred to as willingness, is a binary variable. It represents participants' responses to the question: `Would you authorize us to send you one or more communications about sustainability topics using the Prolific messaging system? The authorization is optional and at your discretion'. Possible answers are `Skip', in which case the variable takes value $0$, and `yes', in which case the variable takes value $1$. We thus employ probit models with various specifications. In all three versions, (1), (2) and (3), the coefficient of the treatment variable remains significant and negative. Since the treatment variable takes a value of 0 in the case of MI and 1 in the case of DS, we can conclude that Motivational Interviewing increases the probability that a participant responds positively to the request for authorization to receive further communications. \emph{Concern about environmental issues}, in models (2) and (3), has positive and highly significant coefficients. Individuals who express greater concern for environmental issues are more likely to accept receiving further communications.

\begin{table}[h!]
\caption{Columns (1)-(3) report the results of probit regressions with \emph{Willingness to receive costly information} as the dependent variable; Columns (4)-(6) report the results of ordered probit regressions with \emph{Self-assessment of satisfaction} as the dependent variable.\label{tab:regression}}
    \resizebox{\textwidth}{!}{%
\begin{tabular}{l|ccc|ccc} \hline
Dep. Var & \multicolumn{3}{c|}{Willingness to receive costly info} & \multicolumn{3}{c}{Self-assessment of satisfaction} \\ [1ex]
 & (1) & (2) & (3) & (4) & (5) & (6) \\ [1ex]\hline
 &  &  & &  &  &  \\
Treatment & -0.126** & -0.130** & -0.113* & -0.0956* & -0.0985* & -0.103* \\
 & (0.0643) & (0.0644) & (0.0652) & (0.0565) & (0.0566) & (0.0570) \\
Concern env. &  & 0.123*** & 0.123*** &  & 0.118*** & 0.136*** \\
 &  & (0.0333) & (0.0344)&  & (0.0307) & (0.0319) \\
Age &  &  & 0.0607***&  &  & 0.0217 \\
 &  &  & (0.0157)&  &  & (0.0138) \\
Age$^2$ &  &  & -0.000715***&  &  & -0.000268 \\
 &  &  & (0.000187)&  &  & (0.000167) \\
Education &  &  & 0.0239 &  &  & -0.0749**\\
 &  &  & (0.0351)&  &  & (0.0316) \\
Income &  &  & -0.00586&  &  & -0.00577 \\
 &  &  & (0.00863)&  &  & (0.00736) \\
Gender &  &  & 0.0795&  &  & 0.0560 \\
 &  &  & (0.0629)&  &  & (0.0555) \\
Constant & 0.396*** & -0.0745 & -1.304*** &  &  &\\
 & (0.0454) & (0.135) & (0.335) &  &  &\\
/cut1&  &  & & -1.077*** & -0.631*** & -0.392 \\
&  &  & & (0.0482) & (0.125) & (0.287) \\
/cut2 &  &  && 0.462*** & 0.918*** & 1.168*** \\
 &  &  && (0.0431) & (0.126) & (0.289) \\
 &  &  & &  &  & \\
 \hline
 Obs. & 1588 & 1588 & 1570& 1,588 & 1,588 & 1,570 \\
 Pseudo R2 & 0.0019 & 0.0083 &0.0172&0.0009 &0.0060 &0.0096 \\ \hline
\multicolumn{7}{c}{\footnotesize Robust standard errors in parentheses, *** p$<$0.01, ** p$<$0.05, * p$<$0.1}
\end{tabular}
}

\end{table}

\emph{Self-assessment of satisfaction}, hereafter referred to as satisfaction, is measured using a smiley rating scale with five options, converted into values ranging from 0 to 4. As evident in Figure 5, the subplot on the right, the satisfaction variable appears particularly imbalanced. Indeed, responses 0, 1, and 2, corresponding to negative and neutral responses, have a markedly lower frequency compared to responses 3 and 4, corresponding to positive responses. To address estimation issues, such as unstable parameter estimates and inflated standard errors, we merge responses (0), (1), and (2) into a single category. The new variable, termed satisfaction pooled, will thus be an ordinal discrete variable with 3 values, (0), (1), and (2). For analysis, we employ an ordered probit with various specifications. Consistent with previous findings, we find that Motivational Interviewing treatment increases the probability that participants express a high degree of satisfaction, in models (4), (5), and (6). Similarly, participants expressing concern about environmental issues are more likely to express a high degree of satisfaction, in models (5) and (6).

As we have seen in the main analysis, the positive impact of DS on learning, expected in the preregistration, is not significant. We now attempt to isolate the direct effects of the treatment on learning from the indirect effects. As observed in the main analysis, MI has a positive and significant effect on interest, and in the exploratory analysis we also find that MI has a positive and significant effect on the time participants spend with the chatbot. Furthermore, the Spearman correlation test reveals that Learning has a positive correlation with both \emph{Time taken} and \emph{Self-assessment of interest}, at any level of significance. We are inclined to interpret \emph{Self-assessment of interest} as a qualitative effect and \emph{Time taken} as a quantitative effect of the treatment. Consequently, MI leads to increased interest and time spent with respect to DS, thus indirectly enhancing learning. To isolate the direct effects of the treatment, we regress \emph{Self-assessment of learning} on the treatment, \emph{Time taken}, and \emph{Self-assessment of interest}, while also controlling for the available socioeconomic variables. To avoid endogeneity issues with the covariates \emph{Time Taken} and \emph{Self-assessment of interest}, an Extended Ordered Probit regression is employed. The total number of words written and the mean number of words per round are used as instruments for \emph{Time Taken} and \emph{Self-assessment of interest}, respectively. 
The results of the Extended Ordered Probit regression are presented in Appendix \ref{appendix}. This analysis suggests that, in line with our expectation, DS does have a positive impact on learning with respect to MI, which is however detected only after controlling for \emph{Time Taken} and \emph{Self-assessment of interest}.

\section{Conclusions}\label{section:conclusions}
This research investigated the effectiveness of communication styles employed by a chatbot designed to conversate with users about sustainable development goals (SDGs). All outcome variables are self-reported measures in a brief questionnaire at the end of the conversation. Our findings suggest that motivational interviewing (MI) significantly increases both engagement and interest in sustainability with respect to directing style (DS). At the same time, no statistically significant difference was observed between MI and DS regarding learning. Therefore, applying MI to environmental issues represents a promising avenue for promoting sustainable behaviors and decision-making. By focusing on individual motivation and resolving ambivalence, MI can be a crucial tool in raising awareness on some of the most pressing environmental challenges of our time, without losing on the received informational content.

Further research could elucidate the extent by which AI-powered MI influences pro-environmental behaviors and explore its efficacy in diverse contexts. For instance, future studies could investigate the application of MI to chatbots designed to discuss about specific sustainability-related subjects, such as energy conservation, waste reduction, or biodiversity protection. Also, AI-powered MI could be explored in application to other societally relevant issues, such as adherence to vaccination campaigns, adoption of healthy lifestyles, attitudes towards immigrants, or gender differences. Another interesting route of research could examine the long-term effects of MI interventions on individuals' attitudes, behaviors, and decision-making processes. Longitudinal studies could provide valuable insights into the durability and persistence of MI-induced changes in pro-environmental behaviors.

Future research could also explore the potential of other communication styles when mediated by chatbot conversations. For example, the efficacy of narrative-based approaches could be investigated \citep{hinyard2007using,richter2019storytelling}, as well as other changes along different dimensions of communication styles \citep{de2009content,de2013communication}. By identifying the most effective communication strategies for different contexts and target audiences, researchers can contribute to the development of tailored interventions that maximize the impact of sustainability messaging.

AI-powered conversational chatbots hold immense potential in amplifying the reach and impact of MI interventions and, more in general, communication interventions. Chatbots can indeed serve as a scalable and cost-effective platform for delivering tailored MI support. This is particularly significant considering the often high costs associated with implementing traditional in-person MI, which can limit accessibility. Future research could focus on optimizing chatbot design, improving natural language processing capabilities, and enhancing the personalization of MI interventions delivered through chatbots. By leveraging the power of AI, researchers can work towards creating more engaging, interactive, and effective tools for promoting sustainable behaviors on a global scale.

\section*{Aknowledgments}

We gratefully acknowledge financial support from the Italian Ministry of Education, University and Research (MIUR) through the PRIN project Co.S.Mo.Pro.Be. ~`Cognition, Social Motives and Prosocial Behavior' (grant n.~20178293XT), and from the European Union - NextGenerationEU through the project ECoHeTE `Effective Communication for Healthcare: Theory and Evidence'.

\section*{Declarations}
E.B., L.B., and E.V. designed research, performed research, and wrote the paper.
\\

\noindent
\textbf{Declaration of generative AI and AI-assisted technologies in the writing process}: During the preparation of this work the authors used ChatGPT for grammar checking. After using this tool/service, the authors reviewed and edited the content as needed and take full responsibility for the content of the publication.\\

\noindent
\textbf{Competing Interests}: The authors declare no competing interests.
\\

\noindent
\textbf{Ethical approval}: Ethical Committee of the University of Florence, approval number: 288/2023.

\bibliographystyle{elsarticle-harv}

\bibliography{DSGs_chatbot}
\newpage
\appendix
\section{Appendix}\label{appendix}
\begin{table}[ht]
\caption{The subtables report the results of an ordered probit regression having endogenous covariates and with \emph{Self-assessment of learning} as dependent variable.}
    \centering
    \resizebox{0.825\textwidth}{!}{
\begin{tabular}{|l|c|c l c}
             \multicolumn{2}{c}{Principal regression} & &\multicolumn{2}{c}{IV regression} \\
\cline{1-2} \cline{4-5}
Learn & Ordered probit & &\multicolumn{1}{|l|}{Interest}& \multicolumn{1}{|c|}{Ordered probit} \\
\cline{1-2}  \cline{4-5}
 &   & & \multicolumn{1}{|c|}{}&\multicolumn{1}{|c|}{}\\
Treatment & 0.0685** &&\multicolumn{1}{|l|}{Words per round} &\multicolumn{1}{|c|}{0.0116***}\\
 & (0.0324)  &&\multicolumn{1}{|c|}{} &\multicolumn{1}{|c|}{(0.00445)}\\
Concern env. &-0.0248   &&\multicolumn{1}{|l|}{/cut1} &\multicolumn{1}{|c|}{-1.477}\\
 & (.0175) &&\multicolumn{1}{|l|}{/cut2} &\multicolumn{1}{|c|}{-1.127}\\
Age & -0.0115 && \multicolumn{1}{|l|}{/cut3}&\multicolumn{1}{|c|}{-0.705}\\
 &(0.00751) && \multicolumn{1}{|l|}{/cut4}&\multicolumn{1}{|c|}{0.189} \\
Age$^2$ & 0.000130 && \multicolumn{1}{|l|}{/cut5}&\multicolumn{1}{|c|}{1.115} \\
\cline{4-5}
 & (0.00009)  &&  &\\

Education & -0.0176  && &\\
 & (0.0114) && \multicolumn{2}{c}{IV regression}\\  \cline{4-5}
Income &0.000792 && \multicolumn{1}{|l|}{Time taken}&\multicolumn{1}{|c|}{OLS}\\
\cline{4-5}
 & (0.00468) &&\multicolumn{1}{|l|}{}&\multicolumn{1}{|c|}{}\\
Gender & -0.016  && \multicolumn{1}{|l|}{Words user}&\multicolumn{1}{|c|}{3.993***}\\
 & (0.0284)&& \multicolumn{1}{|l|}{}&\multicolumn{1}{|c|}{(0.289)}\\
Time taken & 0.0000442 & & \multicolumn{1}{|l|}{Words per round}&\multicolumn{1}{|c|}{-5.93***}\\
 &(0.000147) && \multicolumn{1}{|l|}{}&\multicolumn{1}{|c|}{(2.158)} \\
 Interest &  & & \multicolumn{1}{|l|}{Const.}&\multicolumn{1}{|c|}{235.0713***}\\
\multicolumn{1}{|r|}{1}   & -0.2 & & \multicolumn{1}{|l|}{}&\multicolumn{1}{|c|}{(10.617)}\\
\cline{4-5}
 & (0.14) & & \multicolumn{1}{l}{}&\multicolumn{1}{c}{}\\
  \multicolumn{1}{|r|}{2}   &-0.347** && & \\
   & (0.168) & & \multicolumn{2}{c}{Correlations}\\
   \cline{4-5}
  \multicolumn{1}{|r|}{3}   & -0.513**&&\multicolumn{1}{|l|}{} &\multicolumn{1}{|c|}{}\\
   &(0.245) & & \multicolumn{1}{|l|}{corr(e.Interest,e.Learn)} &\multicolumn{1}{|c|}{0.883***}\\
  \multicolumn{1}{|r|}{4}   & -0.953*** & & \multicolumn{1}{|l|}{} &\multicolumn{1}{|c|}{(0.0483)}\\
   &(0.32) && \multicolumn{1}{|l|}{corr(e.Time taken,e.Learn)} &\multicolumn{1}{|c|}{0.134***} \\
  \multicolumn{1}{|r|}{5}   &-1.412*** && \multicolumn{1}{|l|}{} &\multicolumn{1}{|c|}{(0.0422)} \\
   &(0.423) & & \multicolumn{1}{|l|}{corr(e.Time taken,e.Interest))} &\multicolumn{1}{|c|}{0.168***} \\
 /cut1 &-2.4 & & \multicolumn{1}{|l|}{} &\multicolumn{1}{|c|}{(0.0277)}\\
 \cline{4-5}
 /cut2 & -2.095 && \multicolumn{1}{l}{} &\multicolumn{1}{c}{}\\
 /cut3 &-1.725 & & \multicolumn{1}{l}{} &\multicolumn{1}{c}{}\\
 /cut4 &-1.216 & & \multicolumn{1}{l}{} &\multicolumn{1}{c}{}\\
 \cline{4-5}
 /cut5 & -0.59& & \multicolumn{1}{|l|}{Obs.} &\multicolumn{1}{|c|}{1536}\\
\cline{1-2} \cline{4-5}
\multicolumn{5}{c}{\footnotesize Robust standard errors in parentheses, *** p$<$0.01, ** p$<$0.05, * p$<$0.1}
\end{tabular}
}

\label{tab:regressione_appendice_strumentale}

\end{table}

\begin{table}
\caption{In columns (1)-(3), the results of ordered probit with \emph{Self-assessment of interest} as the dependent variable; In columns (4)-(6), the results of OLS with \emph{Engagement} as the dependent variable.}
\resizebox{\textwidth}{!}{%
\begin{tabular}{l|ccc|ccc}
\hline
Dep. Var.& \multicolumn{3}{c|}{Interest} & \multicolumn{3}{c}{Engagement} \\
 & (1) & (2) & (3) & (4) & (5) & (6) \\
 \hline
 &  &  &  &  &  &  \\
Treatment & -0.0965* & -0.103** & -0.108** & -22.54*** & -22.57*** & -22.47*** \\
 & (0.0524) & (0.0525) & (0.0528) & (2.374) & (2.371) & (2.397) \\
Concern env. &  & 0.259*** & 0.252*** &  & 1.537 & 1.293 \\
 &  & (0.0307) & (0.0311) &  & (1.351) & (1.383) \\
Age &  &  & 0.0325** &  &  & -0.744 \\
 &  &  & (0.0129) &  &  & (0.645) \\
Age$^2$ &  &  & -0.000362** &  &  & 0.0125 \\
 &  &  & (0.000154) &  &  & (0.00804) \\
Education &  &  & 0.0130 &  &  & 0.964 \\
 &  &  & (0.0210) &  &  & (0.941) \\
Income &  &  & -0.0125 &  &  & -0.152 \\
 &  &  & (0.00888) &  &  & (0.413) \\
Gender &  &  & 0.125** &  &  & -4.463* \\
 &  &  & (0.0524) &  &  & (2.336) \\
/cut1 & -1.606*** & -0.659*** & -0.000632 &  &  &  \\
 & (0.0578) & (0.125) & (0.269) &  &  &  \\
/cut2 & -1.245*** & -0.286** & 0.374 &  &  &  \\
 & (0.0494) & (0.123) & (0.269) &  &  &  \\
/cut3 & -0.830*** & 0.143 & 0.807*** &  &  &  \\
 & (0.0441) & (0.123) & (0.269) &  &  &  \\
/cut4 & 0.0665 & 1.066*** & 1.737*** &  &  &  \\
 & (0.0408) & (0.126) & (0.272) &  &  &  \\
/cut5 & 0.985*** & 2.012*** & 2.688*** &  &  &  \\
 & (0.0461) & (0.132) & (0.276) &  &  &  \\
Constant &  &  &  & 56.79*** & 50.87*** & 60.06*** \\
 &  &  &  & (1.897) & (5.664) & (13.73) \\
 &  &  &  &  &  &  \\ \hline
Observations & 1,588 & 1,588 & 1,570 & 1,588 & 1,588 & 1,570 \\
 R2 &  &  &  & 0.053 & 0.054 & 0.063 \\
 Pseudo R2 & 0.0007 & 0.0182 & 0.0212 &  &  &  \\ \hline
\multicolumn{7}{c}{\footnotesize Robust standard errors in parentheses, *** p$<$0.01, ** p$<$0.05, * p$<$0.1}
\end{tabular}
}
\label{tab:regressione_appendice_interest_engage}
\end{table}

\end{document}